\documentstyle[aps,12pt,epsf,preprint]{revtex}
\def \dm {DQ$^2$RM~}
\def \rt {\rightarrow}
\def \beq {\begin{equation}}
\def \eeq {\end{equation}}
\def \be {\begin{eqnarray}}
\def \ee {\end{eqnarray}}

\def \< {\left <}
\def \> {\right >}
\begin{document}
\preprint{
\vbox{
\halign{&##\hfil\cr
        & PKU-TP-98-31\cr
        & hep-ph/9807533 \cr
        & July 1998 \cr\cr}}
        }
\baselineskip 18pt
\renewcommand{\thesection}{\Roman{section}}
\draft
 
\title{Demonstration of the double $Q^2$-rescaling model}

\author{Hong-an Peng$^{1,2}$, Cong-feng Qiao$^{2}$, Jia-sheng Xu$^{1}$,
and Zhen-min He$^{2,3}$}
\address{\small {\it $1$. Department of Physics, Peking University,
Beijing 100871, P.R. China\\
$2$. CCAST (World Laboratory), P.O.Box 8730, Beijing 100080, P.R. China\\
$3$. Department of Physics, HeBei Teachers' University, Shijiazhuang,
050016, P.R. China}}

\maketitle
\begin{abstract}
In this paper we have demonstrated the double $Q^2$-rescaling model
(DQ$^2$RM) of parton distribution functions of nucleon bounded in nucleus.
With different $x$-region of $l$-A deep inelastic scattering process we
take different approach: in high $x$-region ($0.1\le x\le 0.7$) we use the
distorted QCD vacuum model which resulted from topologically
multi-connected domain vacuum structure of nucleus; in low $x$-region
($10^{-4}\le x\le10^{-3}$) we adopt the Glauber (Mueller) multi-scattering
formula for gluon coherently rescattering in nucleus. From these two
approach we justified the rescaling parton distribution functions in bound
nucleon are in agreement well with those we got from DQ$^2$RM, thus the
validity for this phenomenologically model are demonstrated.

\end{abstract}


\section{Introduction}

Deeply understanding the parton distribution functions (PDF) of the
nucleon bounded in nucleus are very important both for studying particle
and high-energy nuclear physics. Since the middle of eighties many
important progress has been achieved both from experimental and
theoretical studies \cite{t1}. It has revealed experimentally that the PDF
of a bound nucleon has remarkable difference from that of a free one: the
EMC effect in high $x$-region ($0.1\le x \le 0.7$) \cite{t2} and the
nuclear shadowing effect \cite{t3} in low $x$-region ($x\le 10^2$) are the
most important examples. Of the EMC effect, many models have been proposed
\cite{t1,t4}, all of them could more or less explain the relevant data,
but for the nuclear shadowing effect only the constituent quark
Q$^2$-rescaling model and the extended $x$-rescaling model can make out
satisfactory explanations after introducing some shadowing parameters
\cite{t1,t5}. As far as we know none of these models could describe the
PDF of a bound nucleon which cover both the
available kinetic x and $Q^2$ range and could explained well all the data
from high energy $l$-A and h-A basic processes. Of course building such a
model is very difficult, but is also very needed. We have tried to take
this task and proposed from phenomenology the double $Q^2$-rescaling model
(DQ$^2$RS) \cite{t6}, from which the calculated results agree rather well
with experimentally relevant data.

In this paper we want to demonstrate our DQ$^2$RM from physical
observations. We take different approach at different $x$-region of $l$-A
DIS. In high $x$-regions (according EMC effect which means ($0.1\le x \le
0.7$), for fixed $Q^2$, it should correspond relatively low energy region
so the non-perturbative QCD effects would be important in this $x$-region.
We observe the QCD vacuum around the bound nucleon is modified
topologically
by multi-connected domain vacuum structure of the nucleus. We further
use soliton bag model of Lee $et~al$ \cite{t7} and find the radius of a
nucleon bounded in nucleus with nucleon number A is larger than that of a
free one, $r_A > r_N$ \cite{t8}. According to uncertainty principle, the
PDF of a bound nucleon would then be soften in comparison with that of a
free one and shift the PDF to small $x$ side, which are consistent with
those from DQ$^2$RM, thus we demonstrated our model at high $x$-region. In
low  $x$-region which covered by available data ($10^{-4}\le x \le
10^{-3}$), it is well known the gluon distribution predominated in PDF of
a nucleus and the coherent interaction of gluons from neighbour nucleons
with same impact parameter in a nucleus becomes important. Following the
method of Levin $et~al$ \cite{t9}, we adopt the Glauber (Mueller)
\cite{t10} multi-scattering formula for calculating contributions of gluon
coherently rescattering in nucleus, thus we get the gluon distribution
function of the nucleon bounded in any nucleus (A) and find it agree well
with those from DQ$^2$RM.

The organic of this paper is as follows: in sec. II we recapitulate the
DQ$^2$RM and its relatives. A detail demonstration on DQ$^2$RM both for
high and low $x$-region will be given respectively in sec.III and sec. IV.
The last sec V is the conclusion and some comments.

\section{Recapitulation of DQ$^2$RM}

Since nuclei are formed from weakly bounded nucleons, one could expected
the basic properties of the nucleon in nucleus, including the behavior of
parton distribution, are not far from that of the free nucleon. All
nuclear effects revealed in $l$-A DIS should be attributed to this nuclear
environment which comes from the weakly bounding mechanism. According to
this observation and informations from relevant phenomenology, we proposed
in \cite{t6} the following relations of PDF between bounded and free
nucleon:
\begin{eqnarray} 
\label{e1}
q^{A}_{V_i}(x,Q^2)=q^{N}_{V_i}(x, \xi^A_V Q^2),~~(i=u,d,s)\nonumber\\
q^{A}_{S_i}(x,Q^2)=\bar{q}^{A}_{S_i}(x,Q^2)=
q^{N}_{S_i}(x, \xi^A_S Q^2),\nonumber\\
g^A(x, Q^2)=g^N(x, \xi_g^A Q^2),
\end{eqnarray} 
where rescaling parameter's $\xi^A_\alpha (\alpha=V, S, g)$ are depending
on and very slowly increased with nucleon number A.

Since it must respect total momentum conservation,
\begin{eqnarray} 
\label{e2}
\int\limits^1_0 x dx \left\{ \sum \limits_i\large[
q^{A}_{V_i}(x, Q^2)+ q^{A}_{S_i}(x, Q^2)+ \bar{q}^{A}_{S_i}(x, Q^2)\large
] + g^A(x, Q^2)\right\}=1.
\end{eqnarray} 
Only two among $\xi^A_\alpha (\alpha =V, S, g)$ are independent, so this
model is called the double $Q^2$-rescaling model.

At present the statistics and the precision of experimental data on
relevant processes are insufficient to test very precisely theoretical
calculation, so in our calculation only the leading order approximation on
the formulas of hard subprocesses and parton distribution functions are
considered. Thus in DQ$^2$RM the structure function and gluon distribution
function of bound nucleon respectively is
\be
\label{e3}
F_2^A(x, Q^2)=\sum \limits_i e^2_i x\left\{
q^{N}_{V_i}(x, \xi^A_VQ^2)+ q^{N}_{S_i}(x, \xi^A_S Q^2)+
\bar{q}^{N}_{S_i}(x, \xi^A_S Q^2)\right\}
\end{eqnarray} 
\be
\label{e4}
x g^A(x, Q^2) =x g^N(x, \xi^A_g Q^2).
\end{eqnarray} 
$F_2^A(x, Q^2)$ denotes the average nucleon structure function of an ideal
nucleus with $N=Z=\frac{1}{2}A$.

The nuclear effect of nucleon structure functioned and the gluon
distribution may be represented by following ratios
\be
\label{e5}
R^{A/D}(x, Q^2) = F_2^A(x, Q^2)/ F_2^D(x, Q^2),
\end{eqnarray} 
\be
\label{e6}
R^{A/D}_g(x, Q^2) = g^A (x, Q^2)/ g^D(x, Q^2),
\end{eqnarray} 

For the Drell-Yan process in p-A collision the nuclear effect is
represented by the ratio 
\be
\label{e7}
T^{A/N}=\int\frac{d^2 {\sigma^{p-A}}(x, x_t, Q^2)}{dx dx_t} dx/\int
\frac{d^2 {\sigma^{p-N}}(x, x_t, Q^2)}{dx dx_t} dx,
\end{eqnarray} 
where
\be
\label{e8}
\frac{d^2 {\sigma^{p-A(N)}}(x, x_t, Q^2)}{dx dx_t}
=\frac{1}{3}\frac{4\pi\alpha^2}{3 m^2_{l\bar l}}\sum\limits_i e^2_i 
\left \{\left [ q^{p}_{V_i}(x, Q^2)+ q^{p}_{S_i}(x, Q^2) \right ]
\bar{q}^{A(N)}_{S_i}(x_t, Q^2)\right.\nonumber\\
+ \left. \bar{q}^{p}_{S_i}(x, Q^2) \left[ q^{A(N)}_{V_i}(x_t, Q^2)
+ q^{A(N)}_{S_i}(x_t, Q^2) \right ]\right \},
\end{eqnarray} 

Three parameters $\xi^A_\alpha (\alpha=V, S, g)$ in DQ$^2$RM are
determined as follows: firstly we choose suitable 
$\xi^A_V$ and $\xi^A_S$ to explain the nuclear effect in $l$-A DIS process
Eq.(\ref{e5});
secondly we describe p-A Drell-Yan process with Eq.(\ref{e7}) and finally
determine $\xi^A_g$ by using nuclear momentum conservation Eq.(\ref{e2})
and explain the nuclear effect of $\gamma^*$-A electroproduction $J/\psi$
process in Eq.(\ref{e6}).

We adopted the PDF of free nucleon given by Gluck- Reya- Vogt (GRV)
\cite{t11} because this formalism allows to be evoluted inversely to
$Q^2\le 1$ GeV$^2$, this is quite important point for getting the PDF of
sea quark and gluon of bound nucleon by Eq.(\ref{e1}) at
$\xi^A_S< 1$ and $\xi^A_g< 1$.

The Q$^2$-rescaling parameters $\xi^A_V$, $\xi^A_S$ and
$\xi_g^A$ on nuclei $C^{12}, Ca^{40}, Fe^{56}$ and $Sn^{119}$ obtained by
above method from the data of references \cite{t3,t12,t13,t14} are listed
in Table 1\footnote{In ref.\cite{t6} the Q$^2$-rescaling constants of
Sn$^{119}$ which we have fitted are $\xi_V^{Sn}=1.57$, $\xi_S^{Sn}=0.45$
and $\xi_V^{Sn}=0.58$, but after recalculating we find it is better to
take these values shown in Table \ref{tb1} instead.}. It is worth to
note that all of $\xi^A_V$ and large than
unity and increased slowly with A, but all of $\xi^A_S$ and $\xi^A_g$ are
smaller than unity and decreased slowly as A increasing.

Using the parameter listed in Table 1 and relevant Eqs. (1)-(8),
together with GRV parametrization for PDF of free nucleon, the theoretical
predications on $R^{A/D}(x, Q^2), T^{A/N}(x, Q^2)$ and $R_g^{A/D}(x, Q^2)$
for nuclei $C^{12}, Ca^{40}, Fe^{56}$ and $Sn^{119}$ by DQ$^2$RM could be
obtained.

The results calculated from DQ$^2$RM in ref.\cite{t6} as follows: 1. The
curves of $R^{A/D}(x. Q^2)$ on $l$-A DIS processes for nuclei
$C^{12}, Ca^{40}, Fe^{56}$ and $Sn^{119}$ within $Q^2$ range given in
\cite{t3,t12} are separatedly plotted in Fig.1(a-d) where corresponding
experimental data \cite{t3,t12} are shown. 2. The curves of $T^{A/D}(x,
Q^2)$ on p-A Drell-Yan process for nuclei $C^{12}, Ca^{40}$ and $Fe^{56}$ 
within region of $0.025\le x\le 0.30$, $4\le m_{l\bar l}\le 9$ GeV and
$E_{CM}=40$ GeV are separately plotted in Fig.2(a-c), where corresponding
experimental data \cite{t13} are shown. 3. The curves of $R_g^{Sn/C}(x,
Q^2)$ (the ratio of gluon distribution function of nucleus Sn to that of
nucleus C) is plotted in Fig.3 where the experimental data \cite{t14} of
the ratio measured in $\gamma^* +A\rightarrow J/\psi + X$ process are
shown.

For comparison we have also plotted in Fig.1-3 the corresponding curves
which calculated from original single Q$^2$-rescaling model (as pointed by
Li $et~al$ in \cite{t15}. This model with a common Q$^2$-rescaling
parameter $\xi(A)$ for valance quark, sea quark and gluon, leads to
non-conservation of nuclear momentum).

From above illustration, we can see that by using DQ$^2$RM all calculated
ratios, which representing the nuclear effects on PDF of bound nucleon,
agree rather well with that from relevant experimental data, Since the
essential elements of this model are embodied in Table 1. we shall
demonstrate it in following sections.

\section{Demonstration for DQ$^2$RM in high $x$-region}

In $l$-A(N) DIS process the Bjorken variable $x=\frac{Q^2}{\hat{S}+Q^2}$,
where $\hat{S}$ is the square of center-mass energy of $\gamma^*$ and
scattered nucleon. For fixed $Q^2$, the higher x region correspond to
lower energy region. Thus in high $x$-region ($0.1\le x\le 0.7$) the
non-perturbative QCD effects should be important and the behavior of PDF
of the nucleon should depends on space-time property of hadron
environment. In the following we will use the soliton bag model of T.D.
Lee \cite{t7} to justify the QCD vacuum structure of a bound nucleon is
topologically different from that of a free one (an isolated nucleon). The
consequence of this difference is the radius of bound nucleon would be
large than that of free one. i.e. $r_A > r_N$. Thus, in comparing with
free nucleon, the PDF of bound nucleon shift to its small x end.

comparing QCD with QED and introducing the concept of colour dielectric
constant $\kappa$, Lee found that the vacuum polarization of QCD shows a
colour anti-screening property. On account of this a phenomenological
scalar field $\sigma(x)$ is introduced into Lagrangian ${\cal L}$ of
Lee's model to describe the colour dielectric property of vacuum. This
$\sigma(x)$ is treated as a classical field with an effective potential
shown in Fig.(\ref{f4}). The potential energy $V(\sigma)$ has an absolute
minimum $V(\sigma_{vac})=0$ at $\sigma=\sigma_{vac}$. which represent the
QCD physical vacuum out of the nucleon bag and which correspond to a
complete colour antiscreening with $\kappa=0$. Inside the bag there are
QCD perturbative vacuum around the colour quarks, where the field
$\sigma=0$ and correspond to $\kappa=1$.

In the case of nucleus, since many nucleons densely populated, the
physical QCD vacuum out of any nucleon which bagged by the nucleus is a
multi-connected domain and so it has a different topologically vacuum
structure comparing with that of free one (isolated nucleon) which is a
single-connected domain. In terms of
Lee model which means, for a nucleon with A nucleons, the $\sigma$ field
in the domain occupied by the remainder (A-1) bags has a value $\sigma=0
(\kappa=1)$ instead of $\sigma=\sigma_{vac}(\kappa=0)$. On the average, 
the effective QCD physical vacuum out of bound nucleon bag correspond to
scalar field $\sigma_A$ ( the relevant vacuum colour dielectric constant
$\kappa_A$) which must smaller than $\sigma_{vac}$, the difference
$(\sigma_{vac}-\sigma_{A})$ increase with increasing of A. From Fig.4. one
could see, comparing with isolated bag where the vacuum pressure on its
surface is B and its equilibrium radius is $r_N$, the bound nucleon bags
would bear smaller vacuum pressure $B_A$ ($B_A < B$) and enlarge
equilibrium bag radius $r_A (r_A > r_N)$. To estimate $\Delta B_A$, a
liner relation between $\sigma$ and $\kappa$ assumed as in Lee model,
\be
\label{e9}
\sigma_A=\sigma_{vac}(1-\kappa_A)
\ee
expand $V(\sigma_A)$ closely around $\sigma_{vac}$ and take the first
non-zero term,
\be
\label{e10}
V(\sigma_A)=b(\sigma_A-\sigma_{vac})^2
\ee
where $b > 0$ according to the definition of physical vacuum. Expressing
$\sigma_A$ by Eq.(\ref{e9}) we get
\be
\label{e11}
\Delta B_A = b \sigma^2_{vac} \kappa_A^2 \equiv G \kappa_A^2
\ee

A simple and intuitive method estimating the effective constant $\kappa_A$
is to consider the volume ratio which the nucleus occupied with that of
(A-1) surrounding nucleons. From geometry picture consideration we get
\be
\label{e12}
\kappa_A=\frac{A^\alpha -1}{\frac{3 A^\alpha}{4 \pi \rho_A r^3_A}-1}
\ee
where $\rho_A$ is the nucleon number density in the nucleus A. In order to
avoid overestimating the effect of those bags far away from the considered
nucleon a parameter $\alpha \le 1$ is introduced and we find
$\frac{1}{3}\le \alpha \le \frac{2}{3}$ are reasonable values.

The relation between $\Delta B_A$ and $r_A$ is determined by
Eqs.(\ref{e11}) and (\ref{e12}). On the other hand the equilibrium
condition of the pressures inside and outside of the bag is determined by
the formula of the bag model \cite{t16}
\be
\label{e13}
M= \frac{C}{r} + \frac{4 \pi}{3} r^3 B
\ee
where $c=3 \epsilon -\alpha_c \sum \limits_{i,j} (\sigma_i 
\sigma_j) (\lambda_i^a \lambda_j^a) \not{h} - Z_0$
is constant in our case, from Eq.(\ref{e13}) we can induce to another
relation between $\Delta B_A$ and $r_A$:
\be
\label{e14}
\Delta B_A = \frac{ 3 M_N r_N}{16\pi}\big
(\frac{1}{r_N^4}-\frac{1}{r_A^4})
\ee 
$M_N$ is the mass of a free nucleon. Combining Eqs.(\ref{e11}),
(\ref{e12}) and (\ref{e14}) we can obtain the self-consistent equation for
determining the nucleon bag radius, $r_A$, in the nucleus;
\be
\label{e15}
G\big (\frac{A^\alpha -1}{\frac{3 A^\alpha}{4 \pi \rho_A r^3_A}-1}\big)^2 
=\frac{ 3 M_N r_N}{16\pi}\big (\frac{1}{r_N^4}-\frac{1}{r_A^4} \big )
\ee
since $G\equiv b \sigma^2_{\rm vac} > 0$, Eq.(\ref{e15}) shows $r_A >
r_N$.

After the EMC effect discovered experimentally, Jaffe and Close
$et~al$ \cite{t4} in their $Q^2$-scale change model proposed that the
effective single nucleon structure functions in a nucleus (A) are related
with that in deuteron by
\be
\label{e16}
F^A_2(x, Q^2)=F_2^D(x, \xi_A Q^2)
\ee
where $\xi_A$ is a scale parameter weakly (logarithmically) dependence on
$Q^2$. On the basis of bag model and QCD they further argued that $\xi_A$
is related to $r_A$ in nucleus by
\be
\label{e17}
\xi_A(Q^2)= (\frac{r_A}{r_N})^{2ln(Q^2/\Lambda^2)/ln(\mu^2/\Lambda^2)}.
\ee

By solving Eq.(\ref{e15}) we get the radius $r_A$ for bound nucleon bag.
There are three parameters $r_N$, $G$ and $\alpha$ in Eq.(\ref{e15}) which
can be determined through fitting the $\xi_A$ values of five nuclei He,
Be, Al, Fe and Au, which are  obtained by rescaling experimental data;
thus we get $r_N=0.865$ fm,
$G=0.0362$ GeV fm, $\alpha=0.541$. In calculation we used $\mu^2=0.63~
GeV^2$, $\Lambda=0.250$ GeV. The results calculated for $Q^2=2.5~GeV^2$
are summarized in Table \ref{tb2}. The experimental values of $\xi_A$ are
not so definite. The difference between $\xi_A$--calculated calculated and
$\xi_A$--experimental data for any nucleus A is around $5\%$.

Eq.(\ref{e16}) is a satisfying phenomenologically relation which all the
data for different nucleus on $l$-A DIS process are in accordance with it.

Comparing the parameters $\xi_\alpha^A (\alpha=V, S, g)$ in Table
\ref{tb1} which comes from DQ$^2$RM with above the parameters $\xi_A$
which describe the effect of QCD vacuum distortion in structure function
of bound nucleon, we notice that:
\begin{enumerate}
\item{Both the values of $\xi^A_V$ in Table \ref{tb1} and $\xi^A$ (in
Table \ref{tb2}) are larger than unity and are weakly increasing as the
nucleon number A increased.}
\item{For any nucleus (A fixed), $\xi_V^A$ always larger than
$\xi^A$, the differnece between them are essentially unvaried with A.}
\end{enumerate}
Furthermore, after taking the $\xi_V^A$ and $\xi_S^A$ from Table \ref{tb1}
and using the PDF of leading order of GRV parametrization, from
Eq.(\ref{e3}) we get $F_2^A(x, Q^2)$ for various nucleus. Use
Eq.(\ref{e16}) once more we get again the Q$^2$-rescaling parameters
$\xi^A_{DQ^2RM}$ for structure functions of bound nucleon from DQ$^2$RM.
We find for any nucleons, the difference between $\xi^A_{DQ^2RM}$ 
with $\xi^A$ in Table \ref{tb2} are less than $5\%$. This is just because
both the parameter $\xi^A_S$ and $\xi_g^A$ in \dm are less than unity and
they are decreasing as A increased, as seen in Table \ref{tb1}.

Thus completed our demonstration for the \dm in high $x$-region.

\section{Demonstration for \dm~ in low $x$-region}

From any phenomenologically parameterization scheme for PDF of a nucleon
we know that in low $x$-region, e.g. $x\le 10^{-3}$ or even smaller, the
distribution function of valence quark almost equal zero whereas that of
the gluon and sea quark are increased very rapidly with the decrease of x;
in addition the distribution function of gluon will nearly be 3 to 5 times
as that of sea quark so as first approximation we will neglect their
contributions. 

In discussion of high energy $\gamma^*$-N(A) DIS process at low
$x$-region, it seems more intuitive and convenient to use the space--time
description approach in target rest frame, $\vec{P}_{N(A)}=0$, rather than
as usually use the Feynman-Bjorken infinite momentum $\vec{P}_{N(A)}\rt
\infty$ approach. In $\vec{P}_{N(A)}=0$ frame, the incoming
virtual photon state $\gamma^*$ at a given instant of time, can be
expanded in terms of its (bare) Fock states \cite{t17}:
\be
\label{e18}
|\gamma^*>_{phys}=\psi_\gamma|\gamma^*> + \psi_{q\bar{q}}|q\bar{q}> +
\psi_{q\bar{q},g}|q\bar{q},g>+\cdots.
\ee

In target rest frame when the scattering energy W of the virtual photon is
vary high and its virtuality $Q^2$ is rather large, we could know this
from \cite{t18} and \cite{t19} that in the probability amplitude
$\psi_{q\bar{q}}$ of decaying into a $q\bar{q}$ pair, there is a rather
large component in which the longitudinal momentum fractions of the quark
and antiquark are nearly equal and their transverse  separation $r_t
(r_t (q\bar{q})) \sim {\cal O}(\frac{1}{Q})$. After scattering off the
target, its transverse expansion $\Delta r_\perp$ is very small,
thus this $q\bar{q}$ pair could be treated approximatedly as a colour
dipole in the interaction with
nucleon (nucleus) target. In this case only those gluons which comes from
the target with transverse momentum $k_\perp\sim {\cal O}(Q)$ could probe
and interact with this colour dipole, so now we can use perturbative QCD
to discuss $\gamma^*$-N(A) DIS process.

Based on this physical observation, Levin $et. ~al$. \cite{t9} using the
generalized Glauber (Mueller) \cite{t10} formula for QCD diffractive
coherent multi-scattering discussed how to get the gluon distribution
function of nucleus, $g^A(x,Q^2)$ from that of nucleon, $g^N(x,Q^2)$.
Since it is relevant for our later discussion, we will first give a
schematically description about their approach.

Since the PDF of nucleon are independence of the probe what we taken, in
order to avoid complexity which comes from quark line loop in
$\gamma^*-g$ fusion subprocess in $\gamma^*$-N(A) DIS, Levin $et~al$
supposed a virtual "gluon probe" ${\cal G}$, which in target rest frame
has very high energy and a rather large virtuality $Q^2$. The probe ${\cal
G}$ decays into a gluon pair GG with nearly equal longitudinal momentum
fraction and their transverse separation $r_\perp$ is rather small. The
kinematics of this approach shown in Fig.\ref{f5}. This pair interacts
with a nucleon due to exchange of a gluon ladder diagram, where $l_\perp$
denotes transverse momentum of the gluon in the ladder attached to GG
pair. The Bjorken variable $x=\frac{Q^2}{\hat{s}}$, where
$\sqrt{\hat{s}}=W$ is the e.m. energy of incoming "gluon" ${\cal G}$. The
change in $r_\perp$ during the
passage of GG pair through the nucleus $\Delta r_\perp\propto
R\frac{k_\perp}{E}$, where E denotes the energy of the pair in target rest
system and $R$ is the size of target, the transverse momentum
$k_\perp\propto\frac{1}{r_\perp}$. It is easy shown when W (and hence E)
is very high and $x \ll\frac{1}{2 m R}$, one get $\Delta r_\perp \ll
r_\perp$ and thus the condition for using Glauber (Mueller) approach in
discussing the GG pair interaction with the nucleus is satisfied. Then
as in \cite{t9} the gluon distribution function of nucleus is given by
\be
\label{e19}
x G_A(x,Q^2)=\frac{4}{\pi^2}\int\limits^1_x\frac{dx'}{x'}
\int\limits^\infty_{\frac{4}{Q^2}}
\frac{d r^2_\perp}{r_\perp^4}\int\limits^\infty_0 \frac{d^2 k_\perp}{\pi}
2\big \{ 1- e^{-\frac{1}{2}\sigma_N^{GG}(x',r_\perp^{2})
S(b_\perp^2)}\big \}.
\ee
The term in curly brackets is the total cross section of the interaction
of gluon pair GG with nucleus in the eikonal approach. In double
logarithmic
approximation (DLA) for perturbative QCD, the GG pair cross section with
the nucleon $\sigma_{GG}(x, r_\perp^2)$ can be written (for $N_c=3$) as 
\cite{t20}
\be
\label{e20}
\sigma_{GG}(x,r_\perp^2)=\frac{3 \alpha_s(\frac{4}{r_\perp^2})}{4}\pi^2
r^2_\perp (x g_N(x, \frac{4}{r^2_\perp)}.
\ee
The factor $\frac{1}{r_\perp^4}$ in Eq.(\ref{e19}) comes from the wave
function of the GG pair in probe ${\cal G}$ and $S(b_\perp)$
is the nucleon profile function in nucleus (A) with impact parameter
$b_\perp$
for ${\cal G}$-A eikonal scattering. Usually it takes Gaussian form:
\be
\label{e21}
S(b_\perp)=
\frac{A}{\pi R^2_A} e^{-\frac{b^2_\perp}{R_A^2}}.
\ee
Putting Eqs.(\ref{e20}) and (\ref{e21}) into Eq.(\ref{e19}) and integrate
over $b_\perp$ we obtain in DLA the gluon distribution of bound nucleon:
\be
\label{e22}
x g^A(x, Q^2)\equiv \frac{1}{A} x G^A(x, Q^2)
=\frac{2 R^2_A}{A \pi^2}\int\limits^1_x\frac{dx'}{x'}
\int\limits^{\frac{1}{Q_0^2}}_{\frac{1}{Q^2}}
\frac{dr^2_\perp}{r_\perp^4} \big \{ C + ln(k_G(x',r^2_\perp)) + E_1
(k_G(x',r^2_\perp)) \},
\ee
where C is the Euler constant and $E_1$ is the exponential integral and 
\be
\label{e23}
k_G(x, r^2_\perp)= \frac{3 \alpha_s A \pi r^2_\perp}{2 R_A^2}x g_N(x,
\frac{1}{r^2_\perp})
\ee
If Eq.(\ref{e22}) is expanded in small $k_G$, the first term will
correspond to the usual DGLAP equations in small $x$ region (Born
approximation of Mueller formula) while the other terms coming from GG
pair coherently multi-scattering with nucleus what will take into account
the shadowing corrections. 

For concrete comparison the $x g^A(x, Q^2)$ calculated, from Glauber
(Mueller) approach with those from \dm, we put $Q^2_0 = 0.5$ GeV$^2$ in
Eq.(\ref{e22}) and calculate the $x g^A(x,Q^2)$ in $2 \le Q^2 \le 4$
GeV$^2$, $10^{-4} \le x \le 10^{-3}$ \footnote{From the DLA condition
which required for validity of Eq.(\ref{e22}), our $Q^2$-range seems
too low, but from the available data at such small $x$-region we found its
$Q^2$-range just so low or even lower. On the other hand from the fact
of "precocious $Q^2$-scaling", which revealed experimentally what the
Bjorken $Q^2$-scaling in $\gamma^*$-N DIS process arrived practically at 
much lower $Q^2$ values than those predicted by perturbative QCD, one
may expected, as a first approximation, one could using Eq.(\ref{e22}) at
such low $Q^2$-range to calculate $xg^A(x, Q^2)$}. In this kinematic
region we introduce the quantity:
\be
\label{e24}
\Delta^{A}_{g}\equiv ~~~{\rm M}&{\rm a}&{\rm x} ~~~ 
\big\{ \frac{x g^A (x, Q^2) - x g^A_{DQ^2RM}(x, Q^2)} {x g^A(x,Q^2)} 
\big\},
\vspace{-0.2cm}
\\
10^{-4}\! \le &x& \le\! 10^{-3}
\nonumber \\
\vspace{-0.2cm}
2\le Q^2& &\!\!\le 4~GeV^2 \nonumber 
\ee
where $x g^A(x, Q^2)$ is calculated from Eq.(\ref{e22}), and
$x g^A_{DQ^2RM}(x, Q^2)$ is deduced from Eq.(\ref{e1}) and Table
\ref{tb1}, both of their
input gluon distribution functions are LO of GRV parameterization sets
\cite{t11}. In addition, when we put the $g^A(x, Q^2)$ resulting from
Eq.(\ref{e22}) into Eq.(\ref{e1}), we get another $Q^2$-rescaling
parameter, $\bar{\xi}_g^A$, for gluon distribution function from Glauber
(Mueller) approach and we can compare it with those from Table \ref{tb1}.

In Table \ref{tb3} we list the $\Delta^A_g, \xi^A_g$ and $\bar{\xi}_g^A$
for 
some nuclei, from it we see the $\Delta_g^A$ values for all but H$^1$
and Be$^9$ nuclei\footnote{Notice Eq.(\ref{e22}) is not an evaluation
equation but a formula for getting $x G^A(x, Q^2)$ of nucleus. In
calculating $x g^A(x, Q^2)$ with Eq.(\ref{e22}), we have let the GRV
parameterization set $g^N(x, Q^2)$ of nucleon as input. Since this input
comes from phenomenology, it should have included effects of rescattering,
thus in some sense, this calculating method have some double counting, but
it is only important for nucleon and few light nuclei.} are less
than $5\%$ and all the gluon Q$^2$-rescaling parameters of \dm, $\xi^A_g$
are all very close to those $\bar{\xi}^A_g$, calculated from Glauber
(Mueller) approach. In addition both the $\bar{\xi}^A_g$ or $\xi^A_g$
seems approach its saturation value, $\geq 0.7$ as $A\geq 200$, this is
related to the fact that in $10^{-4}\le x \le 10^{-3}$ region the coherent
length of the GG pair, $l_{GG}\sim\frac{1}{mx}\approx 10^{3}-10^{4}$ fm
are much large than radius of nuclei, $R_A\le 10$ fm and $R_A$ approch
also its saturation value as $A\geq 200$, so the rescattering number of
GG pair in nucleus also approach saturation.

Here we want to point out that we could not expect in a similar simple
manner as above formalism ---Glauber (Mueller) approach in DLA--- getting
good results from calculating the sea quark distribution function of bound
nucleon, $x q_S^A(x,Q^2)$. In other words, we could not take enough
account the effect of nuclear shadowing on sea quark distribution
functions from equations like Eq.(\ref{e19}) since
there are two points in this case:
\begin{enumerate}
\item{in small $x$-region, e.g. $10^{-4}\le x \le 10^{-3}$ sea quark
distribution functions in nucleon are only 
$\frac{1}{5}$ to $\frac{1}{3}$ of that of the gluon.}

\item{The relevant splitting function are $P_{gq}(x)=P_{g \bar{q}} (x)
=\frac{4}{3}\frac{(1+(1+x)^2)}{x}$,
$P_{gq}(x)=6 [ \frac{x}{(1-x)_+} + \frac{1-x}{x} + x(1-x) 
+(\frac{11}{12}-\frac{N_f}{18})\delta (1-x)]$. at small x limit their
ratio approach $\frac{4}{9}$.}
\end{enumerate}
These two factors would cause the corresponding power for sea quark in the
exponential term in curly bracket smaller about one order of magnitude.
Thus with this simple no mixing (between sea quark and gluon) approach the
resulting effect of nuclear shadowing correction on sea quark is
very weak and the resulting $x q^A_S(x, Q^2)$ would be incorrect.

\section{Conclusion and Comments}

We have adopted different approaches at different $x$-regions to
demonstrate
the \dm~ for PDF of bound nucleon. For high $x$-region, $0.1 \le x \le
0.7$ we used the distorted physical vacuum picture and the soliton bag
model to justify the radius of a bound nucleon always larger than that of
free one. This effect causes the soften of PDF in bound nucleon, which
turns to the $Q^2$-rescaling behavior revealed phenomenologically
on structure functions of bounding nucleon. We thus have demonstrated the
validity of \dm in high $x$-region. For low $x$-region $10^{-4}\le x \le
10^{-3}$ according the formalism of Levin $et~al$ in Glauber (Mueller)
coherently multi-scattering approach, we have shown the resultant gluon
distribution function of bound nucleon, $x g^A(x,Q^2)$ agree very well
with those from \dm. Thus the validity for this phenomenological model are
demonstrated.

Of course, both the \dm on its own and all our demonstrations about it are
far from perfect. We want to make following observations about it.

First, although our demonstrations for \dm are physically reasonable, but
it is far from satisfactory. We could not over the whole $x$-region using 
the perturbative QCD strictly and systematically justified this model,
instead we adopted different approaches for different $x$-region. In
particular we could not give a clear and physical argument for middle
$x$-region, $10^{-1}\le x \le 10^{-3}$, where the contributions for
valance quark, sea quark and gluon are comparatively in $l$-A DIS
processes.

Second, although the \dm reflected approximatedly the objective reality
of PDF in bound nucleon, but it is still a sketchy model. Indeed, if we
have a detailed investigation on this model with perturbative QCD, it is
possible to find that these $Q^2$-rescaling parameters, $\xi^A_\alpha
(\alpha=V, S, g )$, are actually not constants, but perhaps like
logarithmic, weakly depend both on $x$ and $Q^2$. 

Last, a very seriously problem is how to evaluate the non-perturbative QCD
effects exerted on bound nucleons causing by the environment of nucleus.
We have explained in Sec.III the EMC effect with the vacuum distored model
in nuclei is just an example, but such non-perturbative QCD effects would
be active over the whole $x$-region on PDF of bound nucleon.

Obviously, to complete settle these problems from genetic QCD
are very difficult.

\begin{center}
{\bf {\large {Acknowledgements}\ }}
\end{center}

This work was supported in part by the National Natural Science Foundation
of China, the Doctoral Program of Institution of Higher Education of China
and  the Natural Science Foundation of HeBei Province.

\begin{table}
\begin{center}
\begin{minipage}{10cm}
\caption[]{\label{tb1}\small
The fitted values of rescaling parameters.}
\end{minipage}
\doublerulesep 2.0pt
\begin{tabular}{|c|c|c|c|c|}
~~~~ & \hspace{-1.7cm}C$^{12}$~ & \hspace{-1.7cm}Ca$^{40}$~ &
\hspace{-1.7cm}Fe$^{56}$ ~ & \hspace{-1.7cm}Sn$^{119}$ ~
\\\hline
~$\xi_V$ ~ & \hspace{-1.7cm}{1.03}~ & \hspace{-1.7cm}1.35~ &
\hspace{-1.7cm}1.14~ & \hspace{-1.7cm}1.49~
\\ \hline
~$\xi_S$ ~ & \hspace{-1.7cm}0.70~ & \hspace{-1.7cm}0.67~ &
\hspace{-1.7cm}0.62~
& \hspace{-1.7cm}0.60~ \\ \hline
~$\xi_G$ ~ & 0.86~~~~~~~~~~~~~~~& 0.81~~~~~~~~~~~~~~~&
0.76\hspace{1.7cm} & 0.74\hspace{1.7cm} \\
\end{tabular}
\end{center}
\end{table}

\begin{table}
\begin{center}
\begin{minipage}{10cm}
\caption[]{\label{tb2}\small $\xi_A$ and $r_A/r_N$ for various nucleus.}
\end{minipage}
\tabcolsep 1cm
\begin{tabular}{ccc} 
 ~Nucleus~  & ~$\xi_A$~ & ~$r_A/r_N$~  \\ \hline
~He$^4$~ & ~1.223~ & ~1.065~\\ \hline
~Be$^9$~ &  ~1.242~& ~1.070~\\ \hline
~C$^{12}$~ & ~1.248~ & ~1.072~\\ \hline
~A$^{27}$~ & ~1.299~ & ~1.085~\\ \hline
~Ca$^{40}$~ & ~1.319~ & ~1.091~\\ \hline
~Fe$^{56}$~ & ~1.352~ & ~1.099~\\ \hline
~Ag$^{107}$~ & ~1.410~ & ~1.114~\\ \hline
~Au$^{197}$~ & ~1.696~ & ~1.180~ \\ 
\end{tabular}
\end{center}
\end{table}   

\begin{table}
\begin{center}
\begin{minipage}{10cm}
\caption[]{\label{tb3}\small
A list of $\Delta_g^A$, $\xi^A_g$ and $\bar{\xi}^A_g$ for various
nucleus.}
\end{minipage}
\doublerulesep 2.0pt
\begin{tabular}{|c|c|c|c|c|c|c|c|c|c|c|c|} 

 ~A~  & ~H$^1$~ & ~Be$^9$~ & ~C$^{12}$~ & ~Al$^{27}$~ & ~Ca$^{40}$~ &
~Fe$^{56}$~ & ~Cu$^64$~ & ~Sn$^{119}$~ & Ze$^{131}$~ & ~Au$^{17}$~ &
~Pb$^{208}$~ \\ \hline 

~$\Delta^A_g(\%)$~ & ~12~ & ~9~ & ~4~ & ~5~ & ~4~ & ~1~ & ~2~ & ~1~ & ~1~
& ~1~ & ~1~ \\ \hline

~$\bar{\xi}^A_g$~ &  ~1~ & ~0.9~ & ~0.87~ & ~0.84~ & ~0.82~ & ~0.78~
& ~0.76~ & ~0.74~ & ~0.73~ & ~0.72~ & ~0.72~ \\ \hline

~$\xi^A_g$~ &  ~1~ &  & ~0.86~ & ~~ & ~0.81~ & ~0.76~
& ~~ & ~0.74~ & ~~ & ~~ & ~~ \\ 
\end{tabular}
\end{center}
\end{table}   

\begin{figure}
\vspace{-1cm}
\epsfxsize=14 cm
\centerline{\epsffile{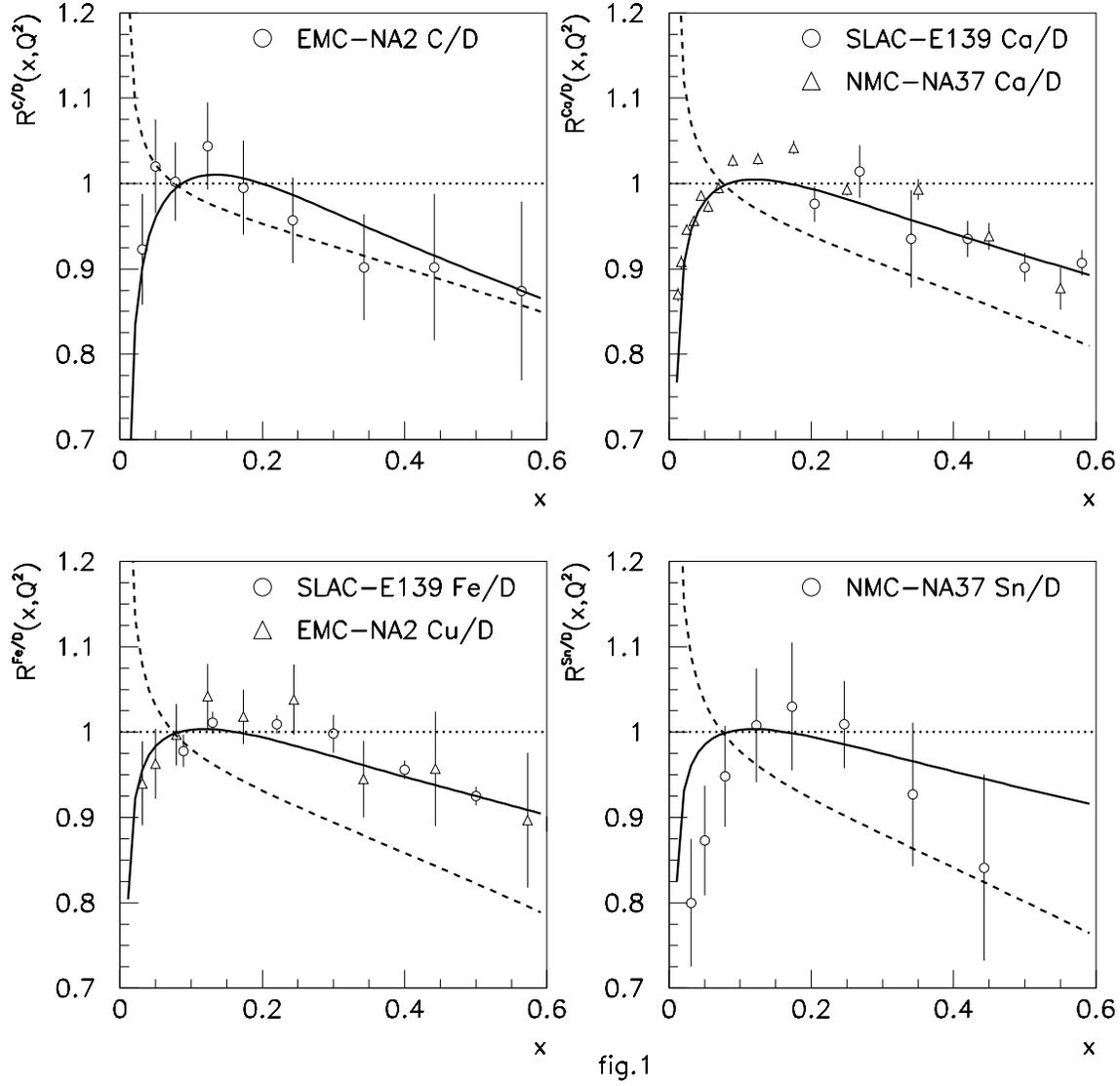}}
\caption[]{\label{f1} The ratio $R^{A/D}(x,Q^2)$  for
nuclei $C^{12}$, $Ca^{40}$, $Fe^{56}$, $Sn^{119}$ versus the momentum
fraction x with the corresponding experimental data \cite{t3,t12}. The
solid lines are the results of our model, the dashed lines are the
results of the old $Q^2-$rescaling model in ref.\cite{t15}.}
\end{figure}

\begin{figure}
\epsfxsize=14 cm
\centerline{\epsffile{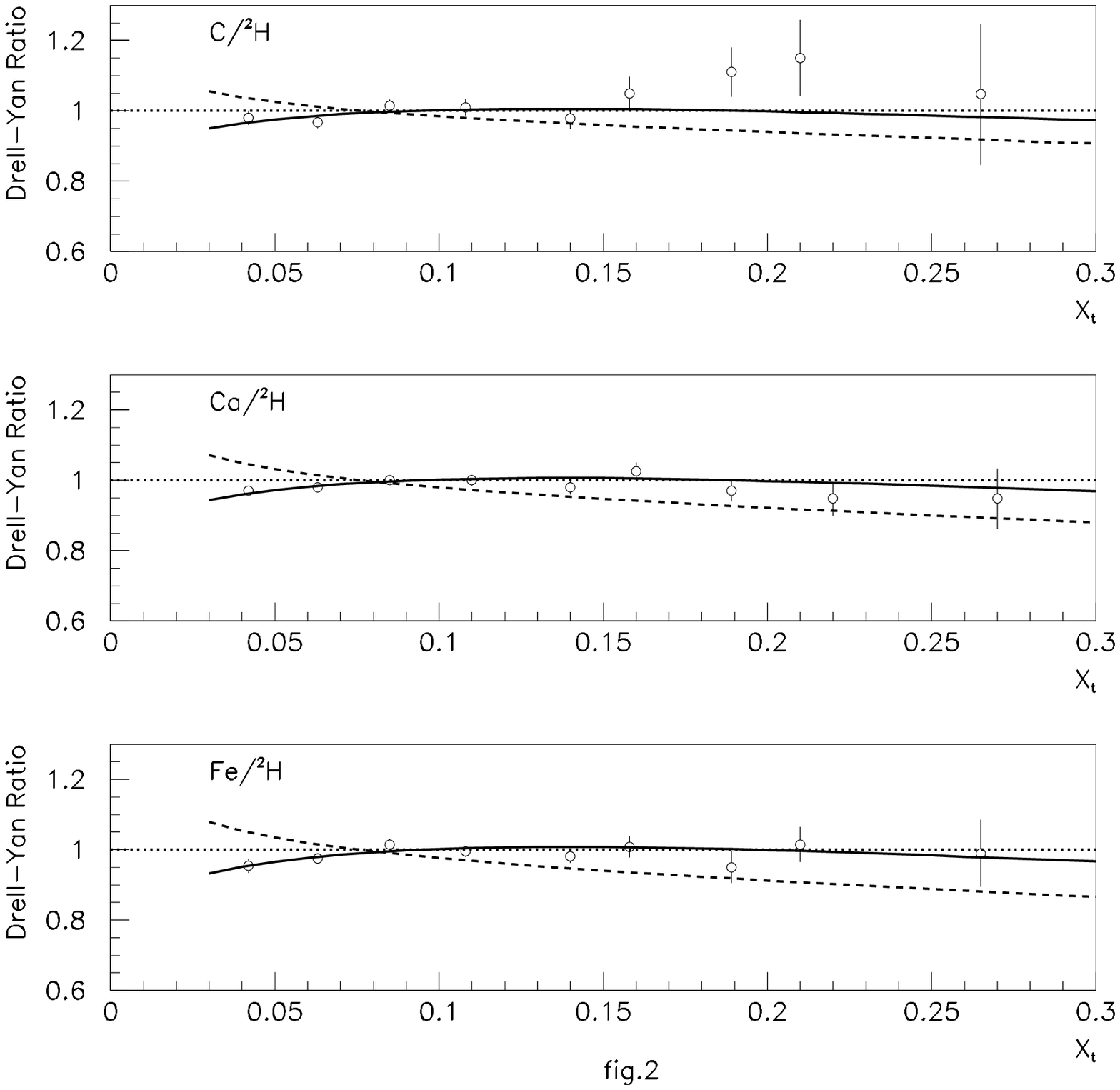}}
\caption[]{\label{f2}
The ratio $T^{A/D}(x_t, Q^2)$ for nuclei $C^{12}$, $Ca^{40}$ and $Fe^{56}$
versus x  within region of $0.025\leq x_t \leq 0.30$, $4\leq M_{l\bar
l} \leq 9GeV$ and $E_{CM}=40GeV$, where corresponding experimental
data \cite{t13}are shown. The meaning of lines is the same as that in
Fig.1.}
\end{figure}
\begin{figure}
\epsfxsize=14 cm
\centerline{\epsffile{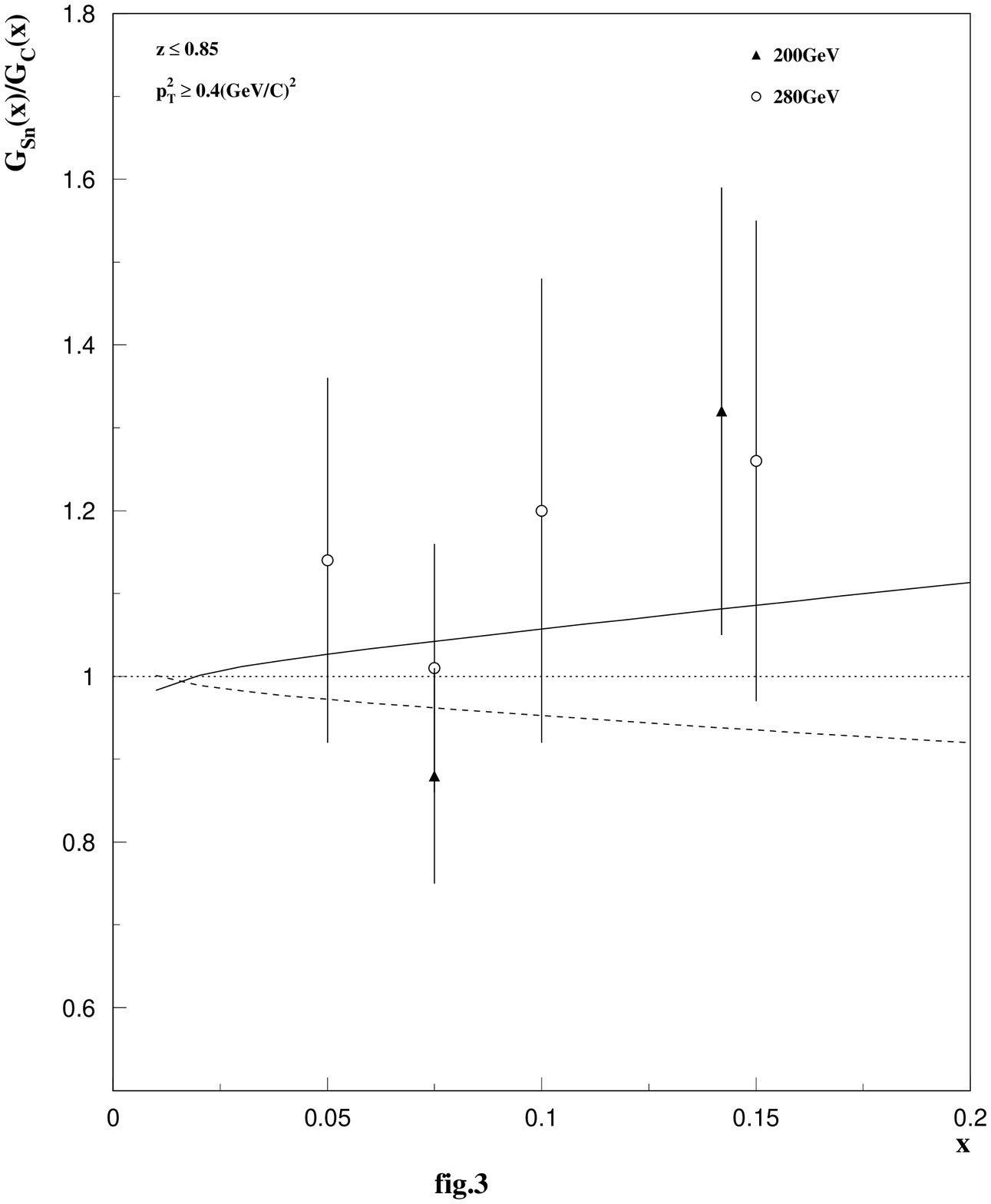}}
\caption[]{\label{f3}
The ratio  $R_{G}^{Sn/C}(x,Q^2)$  versus x, where the experimental
data \cite{t14}are shown. The meaning of lines is the same as that in 
Fig.1.}
\end{figure}

\vskip 6cm
\begin{figure}
\epsfxsize=16 cm
\centerline{\epsffile{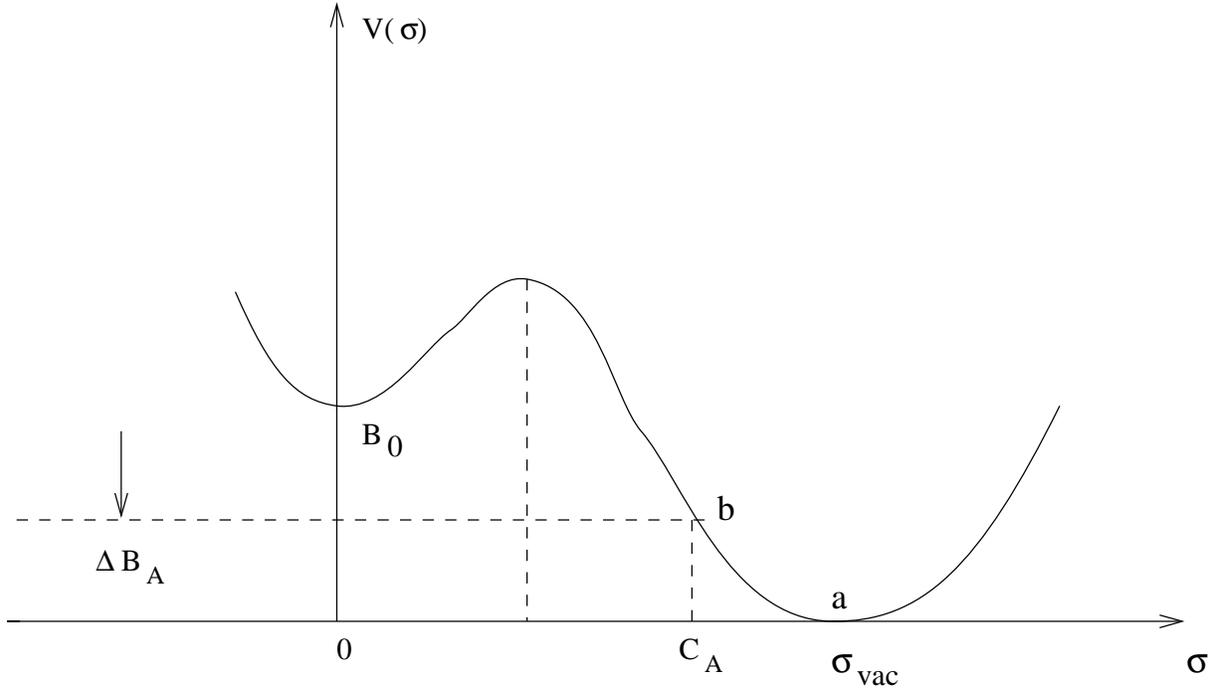}}
\vspace{2cm}
\caption[]{\label{f4} The potential energy of $\sigma$-field and the
change of the bag constant due to the change of $\sigma$-field in the
vacuum.} 
\end{figure}

\begin{figure}
\vspace{-6cm}
\epsfxsize=14cm
\centerline{\epsffile{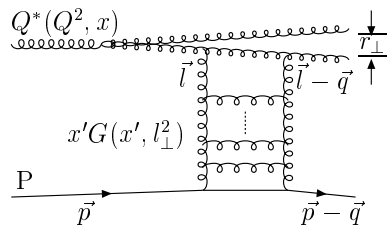}}
\caption[]{\label{f5} Kinematics of the Glauber approach \cite{t9}.}
\end{figure}
\end{document}